\documentclass[12pt]{article}
\usepackage{times}
\usepackage{epsf}
\usepackage{graphicx}

\def\beq{\begin{equation}}
\def\eeq{\end{equation}}
\def\bea{\begin{eqnarray}}
\def\eea{\end{eqnarray}}
\def\ba{\begin{array}}
\def\ea{\end{array}}
\def\bea{\begin{eqnarray}}
\def\eea{\end{eqnarray}}

\begin{document}
\title{Bound state solution of the Schr\"{o}dinger equation for Mie
potential}

\author{Ramazan Sever$^1$\thanks{Corresponding Author:sever@metu.edu.tr },
Cevdet Tezcan$^2$, Mahmut Bucurgat $^3$, \"Ozlem Ye\c{s}ilta\c{s}$^4$ \\[1cm]
$^{1,3}$Department of Physics, Middle East Technical University \\
06531 Ankara, Turkey \\[.5cm]
$^{2}$Faculty of Engineering, Ba\c{s}kent University, Ba\~{g}l{\i}ca
Campus,\\
Ankara, Turkey
\\{\sl $^4$ Turkish Atomic Energy Authority, Istanbul Road, 30 km Kazan 06893, Ankara, Turkey }}

\date{\today}
\maketitle
\normalsize

\begin{abstract}
Exact solution of Schr\"{o}dinger equation for the Mie potential is
obtained for an arbitrary angular momentum. The energy eigenvalues
and the corresponding wavefunctions are calculated by the use of the
Nikiforov-Uvarov method. Wavefunctions are expressed in terms of
Jacobi polynomials. The bound states are calculated numerically for
some values of $\ell$ and $n$ with $n\leq 5$. They are applied to
several diatomic molecules.\vspace{0.3cm}
\end{abstract}

\noindent PACS numbers: 03.65.-w; 03.65.Ge; 12.39.Fd \\[0.2cm]
Keywords: Mie potential, Diatomic molecules, Schr\"{o}dinger
equation, Nikiforov-Uvarov method
%\date{\today}
%\maketitle
%\normalsize

\newpage
\section{Introduction}

\noindent The study of three-dimensional anharmonic oscillators has
raised a considerable amount of interest due to its wide
applications in molecular physics [1-4]. The Morse potential is
commonly used for anharmonic oscillator. However, its wavefunction
does not vanish at the origin. But those of Mie type and
pseudo-harmonic potentials do. On the other hand, the Mie type
potential possess the general features of the true interaction
energy [1]. The Mie type and pseudo-harmonic are two special kinds
of exactly solvable power-law and inverse power-law potentials other
than the Coulombic and harmonic oscillator. The Hamiltonians with
inverse power potential $1/r^{n}$ are studied in different fields.
For instance, path integral, $1/N$ expansion and super-symmetrical
approaches are also studied for this type of potentials [5-8].
Moreover, there are some applications of Mie type potentials such as
equation of bulk metallic glass for
$Zr_{48}Nb_{8}Cu_{14}Ni_{12}Be_{18}$ in solid state physics [9], and
other applications in solid state physics [10], molecular
spectroscopy [11], fluid mechanics [12], the interatomic interaction
potential in molecular physics [13], $1/r^{n}$ potential in three
dimensions [14].

The Nikiforov-Uvarov (NU) method [15] is introduced for the solution
of the hypergeometric type second order differential equations [16]
appeared in the time-independent problems. Recently, NU-method has
been used to solve Schr\"{o}dinger equation for some well known
potentials [17-22], Dirac and Klein-Gordon equations for the Coulomb
and some exponential potentials. In the present work the
Schr\"{o}dinger equation is solved by the NU method for Mie
potential with any value of angular momentum $\ell$.  As an
illustration, energy eigenvalues are computed for $N_{2}$, $CO$,
$NO$ and $CH$ molecules.

The organization of the paper is as follows. In section II, NU
method is briefly introduced. Mie potential calculations are done by
using NU method in section III. Results are discussed in section IV.

\section{The Nikiforov-Uvarov Method}
The NU-method developed by Nikiforov and Uvarov (NU-method) is based
on reducing the second order differential equations (ODEs) to a
generalized equation of hypergeometric type [15]. It is provided us
an analytic solution of Schr\"{o}dinger equation for certain kind of
potentials. It is based on the solutions of general second order
linear differential equation with orthogonal function [16]. The
one-dimensional Schr\"{o}dinger equation is reduced to a generalized
equation of hypergeometric type with an appropriate coordinate
transformation $s=s(r)$. Then, the equation has the form,

\begin{eqnarray}
\psi^{''}(s)+\frac{\tilde{\tau}(s)}{\sigma(s)}\psi^{'}(s)+
\frac{\tilde{\sigma}(s)}{\sigma^{2}(s)}\psi(s)=0
\end{eqnarray}
where $A(s)=\frac{\tilde{\tau}(s)}{\sigma(s)}$ and
$B(s)=\frac{\tilde{\sigma}(s)}{\sigma^{2}(s)}$. In Eq. (1),
$\sigma(s)$ and $\tilde{\sigma} (s)$ are polynomials at most second
degree, and $\tilde{\tau}(s)$ is a polynomial with at most first
degree [15]. The wave function is constructed as a multiple of two
independent parts,

\begin{eqnarray}
\psi(s)=\phi(s) y(s),
\end{eqnarray}
and Eq. (1) becomes

\begin{eqnarray}
\sigma(s)y^{''}(s)+\tau(s)y^{'}(s)+\lambda y(s)=0,
\end{eqnarray}
where

\begin{eqnarray}
\sigma(s)=\pi(s)\frac{d}{ds}(ln \phi(s)),
\end{eqnarray}
and

\begin{eqnarray}
\tau(s)=\tilde{\tau}(s)+2\pi(s).
\end{eqnarray}
$\lambda$ is defined as
\begin{eqnarray}
\lambda_{n}+n\tau^{'}+\frac{[n(n-1)\sigma^{''}]}{2}=0,
n=0,1,2,...
\end{eqnarray}
We determine $\pi(s)$ and $\lambda$ by defining

\begin{eqnarray}
k=\lambda-\pi^{'}(s).
\end{eqnarray}
From the Eqs. (4-5), $\pi(s)$ becomes

\begin{eqnarray}
\pi(s)=(\frac{\sigma^{'}-\tilde{\tau}}{2})\pm
\sqrt{(\frac{\sigma^{'}-\tilde{\tau}}{2})^{2}-\tilde{\sigma}+k\sigma}
\end{eqnarray}
$\pi(s)$ has to be a polynomial of degree at most one in the Eq.
(8), so the expression under the square root must be the square of a
polynomial of first degree [15]. This is simply possible only if its
discriminant is zero. After defining $k$, one can obtain $\pi(s)$,
$\tau(s)$, $\phi(s)$ and $\lambda$. If we look at the Eq. (4) and
the Rodrigues relation

\begin{eqnarray}
y_{n}(s)=\frac{B_{n}}{\rho(s)}\frac{d^{n}}{ds^{n}}[\sigma^{n}(s)\rho(s)],
\end{eqnarray}
where $C_{n}$ is normalizable constant and the weight function
satisfy the relation as

\begin{eqnarray}
\frac{d}{ds}[\sigma(s)\rho(s)]=\tau(s)\rho(s).
\end{eqnarray}
where

\begin{eqnarray}
\frac{\phi^{'}(s)}{\phi(s)}=\frac{\pi(s)}{\sigma(s)}.
\end{eqnarray}

\section{Mie Potential Calculations}
The Mie-type potentials are given by

\begin{equation}
V(r)=\epsilon\left[\frac{k}{\ell-k}\left(\frac{a}{r}\right)^{\ell}-
\frac{\ell}{\ell-k}\left(\frac{a}{r}\right)^{k}\right]\label{eq2}
\end{equation}
where $\epsilon$ is the interaction energy between two atoms in a
solid at $x=a$, and $\ell > k$ is always satisfied.\vspace{0.3cm}

We solve the one-dimensional Mie potential [1-2] with $\ell=2k$
combination. By choosing the special case $k=1$, corresponding to a
Coulombic-type potential with an additional centrifugal potential
barrier, we get the following form:

\begin{equation}
V(r)=V_{0}\left[\frac{1}{2}\left(\frac{a}{r}\right)^{2}-\left(\frac{a}{r}\right)\right];
\;V_{0}=2\epsilon k \label{eq2}
\end{equation}
where $V_{0}$ is the dissociation energy and $a$ is the positive
constant which is strongly repulsive at shorter distances. The
radial part of the Schr\"{o}dinger equation for a diatomic molecule
potential is

\begin{equation}
\left[-\frac{\hbar^{2}}{2\mu}\frac{1}{r^{2}}\frac{d}{dr}\left(r^{2}\frac{d}{dr}\right)+\frac{\ell(\ell+1)\hbar^{2}}{2\mu
r^{2}}+V(r)\right]R_{n\ell}(r)=E_{n\ell}R_{n\ell}(r),\label{eq3}
\end{equation}
where $\mu$ is the reduced mass of the diatomic molecules. $n$
denotes the radial quantum number ($n$ and $\ell$ are named as the
vibration-rotation quantum numbers in molecular chemistry). $r$ is
the internuclear separation. Substituting the explicit form of
$V(r)$, we get

\begin{equation}
\frac{d^{2}R_{n\ell}(r)}{dr^{2}}+\frac{2}{r}\frac{dR_{n\ell}(r)}{dr}
+\frac{2\mu}{\hbar^{2}}\left[E_{n\ell}-V_{0}\left(\frac{1}{2}\left(\frac{a}{r}\right)^{2}-\frac{a}{r}\right)\right]R_{n\ell}(r)
-\frac{\ell(\ell+1)\hbar^{2}}{2\mu r^{2}}R_{n\ell}(r)=0.\label{eq4}
\end{equation}
By defining the following variables:
\begin{eqnarray}
\varepsilon^{2}=E\label{eq5}\\[0.2cm]
\beta=-\frac{2\mu}{\hbar^{2}}V_{0}a\label{eq6}\\[0.2cm]
\gamma=\frac{2\mu}{\hbar^{2}}\left(\frac{1}{2}V_{0}a^{2}+\frac{\ell(\ell+1)\hbar^{2}}{2\mu}\right)\label{eq7}
\end{eqnarray}
The Schr\"{o}dinger equation takes the simple form:

\begin{equation}
\left[\frac{d^{2}}{dr^{2}}+\frac{2}{r}\frac{d}{dr}+\frac{1}{r^{2}}\left(\varepsilon^{2}r^{2}-\beta
r-\gamma\right)\right]R_{n\ell}(r)=0\label{eq8}
\end{equation}
From Eq. (19), it is clear that $\sigma(r)=r$, $\tilde{\tau}=2$,
$\tilde{\sigma}=\varepsilon^{2}r^{2}-\beta r-\gamma$. We find
$\pi(r)$ as

\begin{equation}
\pi(r)=-\frac{1}{2}\pm \left\{
                        \begin{array}{ll}
                          i \varepsilon r+\frac{1}{2} \sqrt{4 \gamma+1}, & \hbox{$k_{1}=- \beta+\varepsilon \sqrt{-1-4\gamma}$ ;}
\\
                          i \varepsilon r-\frac{1}{2} \sqrt{4 \gamma+1}, & \hbox{$k_{2}=- \beta-\varepsilon \sqrt{-1-4\gamma}$.}
                        \end{array}
                      \right.
\end{equation}
and $\tau (r)$ can be written as

\begin{equation}
\tau(r)= \left\{
               \begin{array}{ll}
                 1+2i\varepsilon r+\sqrt{4 \gamma+1}, & \hbox{$k_{1}=- \beta+\varepsilon \sqrt{-1-4\gamma}$;} \\
                 1-2i\varepsilon r+\sqrt{4 \gamma+1}, & \hbox{$k_{2}=- \beta-\varepsilon \sqrt{-1-4\gamma}$.}
               \end{array}
             \right.
\end{equation}
For appropriate solutions $\tau^{'}(r) < 0$ [15], we use $k_{2}=-
\beta-\varepsilon \sqrt{-1-4\gamma}$,
$\pi_{2}=-\frac{1}{2}-i\varepsilon r-\frac{1}{2} \sqrt{1+4\gamma}$
and $\tau(r)=1-2i\varepsilon r+\sqrt{1+4\gamma}$ for obtaining
eigenvalues and eigenfunctions. Using Eqs. (6-7) we find energy
spectrum as

\begin{equation}
E_{n}=-\left(\frac{2\mu V_{0}
a}{\hbar^{2}}\right)^{2}\left[2n+1+\sqrt{1+\frac{8\mu}{\hbar^{2}}
\left(\frac{V_{0}a^{2}}{2}+\frac{\ell
(\ell+1)\hbar^{2}}{2\mu}\right)}\right]^{-2}
\end{equation}
Wavefunction is calculated from Eqs. (9-11).

\begin{equation}
\phi(r)=r^{\frac{1}{2}(-1+\sqrt{1+4\gamma})} e^{-i \varepsilon r}
\end{equation}
and

\begin{equation}
\rho(r)=r^{\sqrt{1+4\gamma}} e^{-2i \varepsilon r}
\end{equation}
and the radial part of the wave function becomes

\begin{equation}
R_{n,\ell}=C_{n}r^{\frac{1}{2}(-1+\sqrt{1+4\gamma})} e^{-i
\varepsilon r} L^{\sqrt{1+4\beta}}_{n}(2i \varepsilon r)
\end{equation}
If we rewrite Eq. (22) in atomic units, we obtain

\begin{equation}
E_{n}=-V^{2}_{0} a^{2}
\left(2n+1+\sqrt{(2\ell+1)^{2}+2V_{0}a^{2}}\right)^{-2}
\end{equation}

\section{Conclusions}
We have investigated the analytical solution of the Schr\"{o}dinger
equation for Mie potential. By using a special case of Mie potential
as $m=2k$ and $k=1$, the problem is reduced to a Coulombic potential
with the additional centrifugal barrier. Energy eigenvalues and
corresponding eigenfunctions are obtained by the NU method. energy
eigenvalues are computed for $N_{2}$, $CO$, $NO$ and $CH$ molecules.
They are listed in Table 1.

\section{Acknowledgements}

This research was partially supported by the Scientific and
Technological Research Council of Turkey.

\newpage
\begin{table}[tbp]
\caption{Calculated energy eigenvalues of the Mie potential for $%
N_{2},$ $CO,$ $NO$ and $CH$ diatomic molecules with different values
of $n$ and $\ell$ in $eV.$The data for potential parameters are
taken from [22].}
\begin{tabular}{llllll}
State $(n)$ & $\ell$ & $N_{2}$ & $CO$ & $NO$ & $CH$ \\
$0$ & $0$ & $0.00271631$ & $0.0254063$ & $0.0205538$ & $0.0416018$
\\
$1$ & $0$ & $0.0810235$ & $0.0756383$ & $0.0611503$ & $0.120556$ \\
& $1$ & $0.0810268$ & $0.0758723$ & $0.0613568$ & $0.122185$ \\
$2$ & $0$ & $0.13411$ & $0.125172$ & $0.101132$ & $0.194768$ \\
& $1$ & $0.134351$ & $0.125403$ & $0.101335$ & $0.1963$ \\
& $2$ & $0.134833$ & $0.125867$ & $0.101742$ & $0.199356$ \\
$3$ & $0$ & $0.186486$ & $0.17402$ & $0.140511$ & $0.264609$ \\
& $1$ & $0.186719$ & $0.174248$ & $0.140712$ & $0.266052$ \\
& $2$ & $0.187194$ & $0.174704$ & $0.141112$ & $0.26893$ \\
& $3$ & $0.187907$ & $0.175387$ & $0.141713$ & $0.273228$ \\
$4$ & $0$ & $0.238152$ & $0.222196$ & $0.1793$ & $0.330417$ \\
& $1$ & $0.238385$ & $0.222421$ & $0.179498$ & $0.331777$ \\
& $2$ & $0.238854$ & $0.22287$ & $0.179892$ & $0.334491$ \\
& $3$ & $0.239557$ & $0.223544$ & $0.180484$ & $0.338544$ \\
& $4$ & $0.240495$ & $0.224442$ & $0.181272$ & $0.343916$ \\
$5$ & $0$ & $0.28913$ & $0.269712$ & $0.217511$ & $0.392495$ \\
& $1$ & $0.289361$ & $0.269933$ & $0.217705$ & $0.393779$ \\
& $2$ & $0.289824$ & $0.270377$ & $0.218094$ & $0.396341$ \\
& $3$ & $0.290518$ & $0.271041$ & $0.218677$ & $0.400166$ \\
& $4$ & $0.291443$ & $0.271927$ & $0.219453$ & $0.405238$ \\
& $5$ & $0.292599$ & $0.273036$ & $0.220424$ & $0.41153$ \\%
\end{tabular}
\end{table}
\end{document}